\documentclass[prl,aps,showpacs,footinbib,twocolumn,final,amsmath]{revtex4-1}
\usepackage{graphics,graphicx,color}

\begin{document}


\title{Magnetic dephasing in mesoscopic spin glasses}

\author{Thibaut Capron$^{1}$, Guillaume Forestier$^{1}$, Angela Perrat-Mabilon$^{2}$, Christophe Peaucelle$^{2}$, 
Tristan Meunier$^{1}$, Christopher B\"{a}uerle$^{1}$, Laurent~P.~L\'{e}vy$^{1,4,5}$, David Carpentier$^{3}$ and 
Laurent Saminadayar$^{1,4,5,\dagger}$}
\affiliation{$^1$Institut N\'{e}el, B.P. 166, 38042 Grenoble Cedex 09, France}
\affiliation{$^2$Institut de Physique Nucl\'{e}aire de Lyon, Universit\'{e} de Lyon, 
Universit\'{e} de Lyon 1, \textsc{cnrs} \& \textsc{in2p3}, 04 rue Enrico Fermi, 69622 Villeurbanne Cedex, France}
\affiliation{$^3$Laboratoire de Physique, \'{E}cole Normale Sup\'{e}rieure de Lyon, 47 all\'{e}e d'Italie, 69007 Lyon, France}
\affiliation{$^4$Universit\'{e} Joseph Fourier, B.P. 53, 38041 Grenoble Cedex 09, France}
\affiliation{$^5$Institut Universitaire de France, 103 boulevard Saint-Michel, 75005 Paris, France}

\date{\today}


\begin{abstract}
We have measured Universal Conductance Fluctuations in the metallic spin glass $Ag:Mn$ as a function of temperature and magnetic field. 
From this measurement, we can access the phase coherence time of the electrons in the spin glass. We show that this phase coherence time
increases with both the inverse of the temperature  and the magnetic field. From this we deduce that decoherence mechanisms are still 
active even deep in the spin glass phase.
\end{abstract}

\maketitle


Spin glasses are one of the most fascinating states of matter. It has attracted the interest of a large community for several decades, as it 
is one of the most fundamental problem in Condensed Matter Physics~\cite{Houches_06_AA}. Spin Glass appears when  magnetic atoms are randomly diluted in a 
non-magnetic metallic host. As the spatial distribution is random, so are the \textsc{rkky} interactions between the spins. This leads to a
frustration between the magnetic moments and their couplings. It is this interplay between disorder and frustration that leads to the 
formation of a \emph{spin glass} below a transition temperature $T_{sg}$. The very nature of the ground state is still 
heavily debated and may consist in  an unconventional state of matter with remarkable behavior~ \cite{Vincent_09_AA}. Transport properties of these 
metallic spin glasses, however, have not been studied thoroughly. In particular, only very few studies, experimental or theoretical, 
have addressed the question of the quantum coherence of the electrons in a metallic spin glass, and it is usually taken for granted that, 
as spins are frozen, inelastic scattering processes due to spins are also frozen and one should recover the coherence time observed in 
a Fermi liquid.

It has been widely recognized that transport measurement can be a powerful tool to probe the quantum coherence in metallic 
systems~\cite{Schopfer_03_AA}, and that this concept can be extended to spin glasses~\cite{Levy_90_AA}. Recently, the interest in 
quantum transport measurements in spin glasses has been even renewed thank to the idea that this type of measurements could give 
access to the structure of the ground state of the system~\cite{Carpentier_08_AA}.

In this Letter, we present measurements of Universal Conductance Fluctuations (\textsc{ucf}s) in a metallic spin glass as a function of
temperature and magnetic field. From this we deduce the phase coherence time of the electrons; we show it increases as the temperature
decreases, in agreement with theoretical predictions. Moreover, from the magnetic field dependence of the decoherence rate, we show
that decoherence mechanisms persist deep in the spin glass phase.
\begin{figure}[t!]
\includegraphics[width=8.5cm]{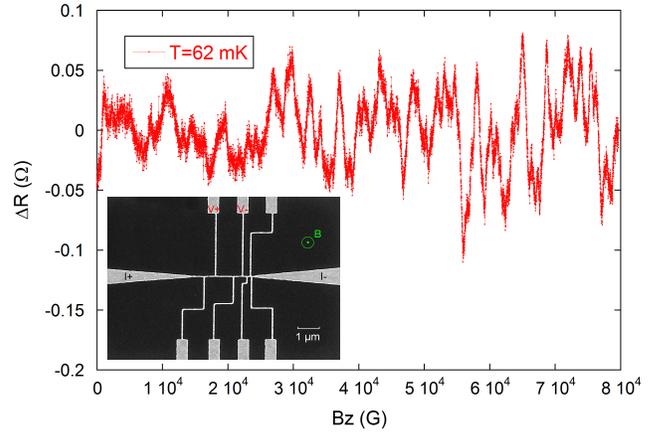}
\caption{(Color online) Magnetoresistance of a spin glass $Ag$ sample implanted with $700\,ppm$ of manganese. The magnetic field
is applied perpendicularly to the sample. Insert: Scanning Electron Microscopy of the sample; current and voltage probes are indicated 
in red.}
\label{Fig1}
\end{figure}

Samples have been fabricated on a silicon/silicon oxide wafer using standard electron-beam lithography on polymethyl-methacrylate
 (\textsc{pmma}) resist. Geometry of the sample consists in a long ($\rm{length}\, L \approx 2\,\mu m$) and thin 
 ($\rm{width}\, w \approx 50\,nm$, $\rm{thickness}\, t \approx 40\,nm$) wire (see inset of figure \ref{Fig1}). Several contacts have 
 been put along the wire, in order to measure the resistance over different lengths and to thermalize properly the electrons 
 along the wire. Silver has been evaporated using a dedicated electron gun evaporator and a $99.9999\%$ purity source without adhesion 
 layer. Electronic properties of pure samples fabricated during the same run have been measured in previous 
 works~\cite{Mallet_06_AA,Capron_11_AA}, leading to the values $l_{e} = 43\, nm$ for the elastic mean free path, 
 $\rho = 2.15\,\mu\Omega\cdot cm$ for the resistivity and $L_{\phi} = 7\,\mu m$ for the phase coherence length at low temperature 
 (below $\approx 50\,mK$). Samples have then been implanted with $Mn^{2+}$ ions of energy $70\,keV$. The energy has been chosen 
 after numerical simulations based on the \textsc{srim} software in order to ensure that ions will end up in the sample following a 
 gaussian distribution whose maximum lies in the middle of the sample thickness. Moreover, this technique of implantation 
 allows to avoid clustering or migration of the $Mn^{2+}$ ions as no further annealing has been performed on the samples. The ion 
 dose, measured \textit{via} the current of the implanter, has been chosen in order to give a final ion concentration in the wire of 
 $700\, ppm$; such a concentration leads to a transition temperature of $T_{sg}\approx 700\, mK$, as has been demonstrated 
 recently by transport measurements~\cite{Capron_11_AA}: it is thus easy to perform measurements well below $T_{sg}$.
\begin{figure}[th!]
\includegraphics[width=8.5cm]{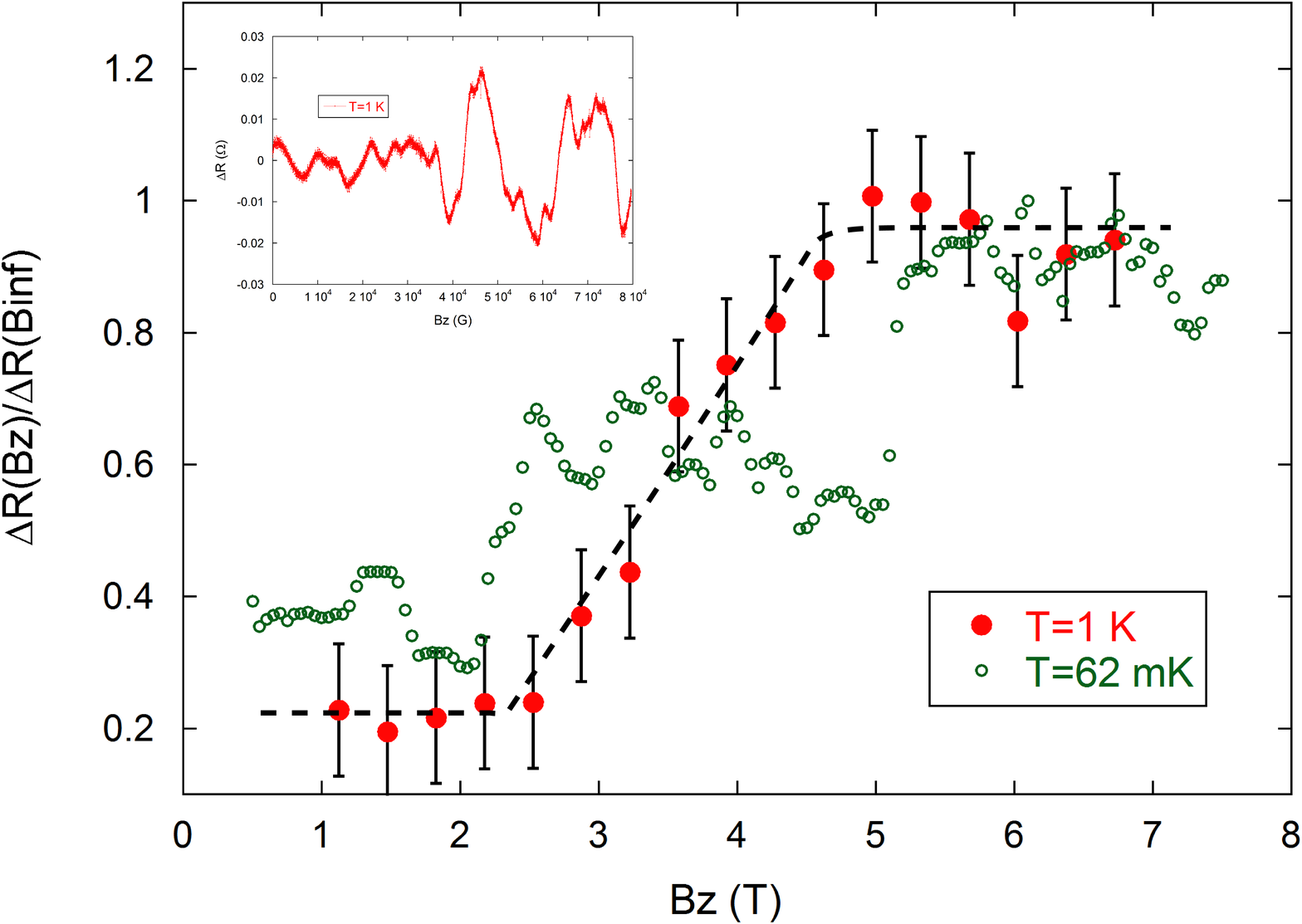}
\caption{(Color online) Relative amplitude of the Universal Conductance Fluctuations as a function of the magnetic 
field. The inset shows that this increase is already clear on the direct magnetoresistance trace.}
\label{Fig2}
\end{figure}
The sample has been cooled down in a dilution fridge whose base temperature $T$ is 
$\approx 50\,mK$ and equipped with a 
superconducting coil of maximum magnetic field $B = 8\, T$. Transport measurements have been 
carried out using an ac lock-in technique 
at a frequency of $11\, Hz$ in a bridge configuration and using a very low current of $250\, nA$, in order to avoid any over heating of the sample. 
Signal is amplified using an ultra-low 
noise homemade voltage amplifier (voltage noise $S_{v} = 500\, pV/\sqrt{Hz}$) at room temperature. 
All the measuring lines connecting
 the sample to the experimental set-up consist in lossy coaxes which ensure a very efficient 
 radio-frequency filtering and thus a good 
 thermalization of the sample~\cite{Zorin_95_AA,Glattli_97_AA,Mandal_11_AA}. At $4.2\, K$, 
 the resistance of our sample is 
 $R \approx 10\,\Omega$.

Universal Conductance Fluctuations are obtained by measuring the magnetoresistance of the sample between $0\, T$ and $8\, T$. 
The field is swept at a rate of $20\, mG/s$ in order to avoid excessive eddy-currents heating of the fridge. After subtraction of the classical magnetoresistance of a spin glass~\cite{Laborde_71_AA,Mydosh_74_AA,Larsen_76_AA}, one clearly sees the \textsc{ucf}s as shown on figure~\ref{Fig1}. These magnetoresistance traces  are decorrelated by a 
temperature cycling well over $T_{sg}$, signaling the coupling of this magnetoresistance to the frozen spins. 
On the other hand it should be stressed that they are completely reproducible over the sweeping direction. 
Moreover, we observe that the maximum magnetic field applied is much larger than the expected critical field, $B\gg B_c = k_{B}T_{sg}/\mu_{B}$ 
with $\mu_{B}$ the Bohr magneton and $k_{B}$ the Boltzmann constant. Such a result is similar to what has been observed in the 
work of L\'{e}vy \textsl{et al.} on a $Cu:Mn$ system~\cite{Levy_90_AA}: \emph{the exact spin configuration 
of a spin glass is preserved after a magnetic cycling to very high magnetic field at a temperature of $\approx T_{sg}/10$}. Within the mean-field phase diagram of spin glasses  two 
critical lines exist related to the freezing of longitudinal component (\textquotedblleft de Almeida-Thouless\textquotedblright~line) and 
transverse component of the spins (\textquotedblleft Gabay-Toulouse\textquotedblright~line). While magnetization measurements usually
probe the longitudinal freezing and observe a freezing field of the order $B_c$, transport measurements are sensitive to the freezing of both
components of the spins. Similarly, this transverse freezing has also been probed in torque experiments as reported in~\cite{Petit:02}.
Remarkably, both experiments signal much stronger freezing field than the expected $B_c$. Whether this finding is compatible with 
the chiral freezing scenario of Kawamura \textsl{et al.}~\cite{Kawamura:01} is an interesting although open question.

We now consider the \emph{amplitude} of these \textsc{ucf}s as a function of the magnetic field; for this purpose, we have plotted on 
figure~\ref{Fig2} the peak-to-peak amplitude (extracted from the inset of the figure) as a function of the 
magnetic field. Two regimes 
are clearly visible: a low field regime (typically below $B \lesssim 2.5\, T$) 
and a high field regime ($B\gtrsim 4.5\, T$) regimes. They differ by the amplitude of the UCFs which
is \emph{$5$ times larger} in the high field regime. 
Such an increase of the amplitude of the \textsc{ucf}s manifests an increase of 
$L_{\phi}$~\cite{Akkermans_07_AA}. More precisely as long as the phase coherence length of the electrons $L_{\phi}$ is
 smaller than both the system length and the thermal length $L_\phi \leq L_T,L$, 
 the amplitude of the UCFs scale as 
 $\Delta G_{\textsc{ucf}s} \propto (L_{\phi}/L)^{3/2}$. 
 In this regime, an increase of the UCF manifests the corresponding  increase of
 dephasing length. On the other hand, for $L_T \leq L_\phi \leq L$ the UCFs amplitude saturates to 
  $\Delta G_{\textsc{ucf}s} \propto (L_{T}/L)$ and becomes independent of $L_\phi$. The results of figure~\ref{Fig2} are thus interpreted as an increase of the
dephasing length $L_\phi$ up to a magnetic field $B \simeq 4.5\, T$, until 
this dephasing length reaches the thermal length $L_\phi \simeq L_T$  or system size 
$L_\phi \simeq L$ for $B\simeq 4.5 T$. 
 Below, we will relate this increase of dephasing length to a change of magnetic
dephasing due to the polarization of the ensemble of magnetic impurities.  
The scale in magnetic field on 
which this polarization occurs is governed by  the typical amplitude  $J_{{\textsc{\footnotesize rkky}}}$ of the couplings
between impurity spins. The magnetization of $Cu:Mn$ bulk spin glass with an $Mn$ concentration in the range $1-10\,\%$ in high field up to $40\,T$ shows a progressive saturation with field~\cite{Smit_79_AA}. Rescaling these data by the impurity concentration we infer a saturating field of $2.5\,T$ for our $700\,ppm$ $Ag:Mn$ sample, compatible with the data represented on figure~\ref{Fig2}.
\begin{figure}[t!]
\includegraphics[width=8.5cm]{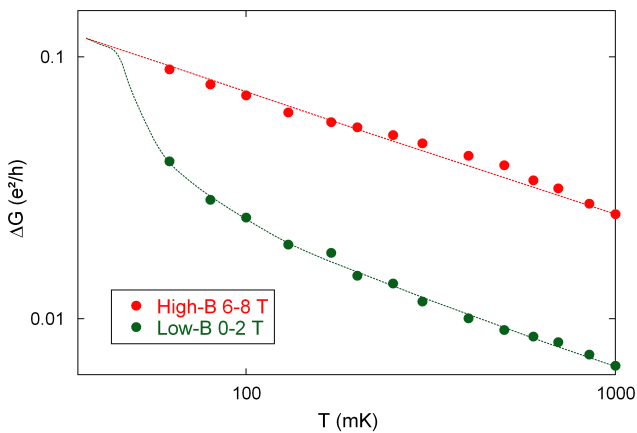}
\caption{
(Color online) Amplitude of the \textsc{ucf}s as a function of temperature, for high field ($B\geq6\,T$) and low field 
($B\leq2\,T$ field). Note that the \textquotedblleft high\textquotedblright~field curve follows nicely a power law, whereas 
the \textquotedblleft low\textquotedblright~field curve tends to join it at very low temperature. Dotted lines are guides for the eyes.}
\label{Fig3}
\end{figure}
We now focus on the low and intermediate field regimes. For the range of parameters explored here, 
the electron-phonons and electron-electron mechanism are subdominant
and 
the dephasing rate $1/\tau_\phi$ is dominated by the 
electron-magnetic impurities processes \cite{Mallet_06_AA}, $1/\tau_{\phi} \simeq 1/\tau_{e-s}$.
We thus focus on the magnetic dephasing rate, and its evolution with magnetic field
and temperature. We
account for the strong couplings between the magnetic impurities through an
effective field description \cite{Boettcher:08}. In this description, the spin $\vec{S}_i$ 
of each magnetic impurity is submitted to a effective magnetic field $\vec{h}_i$ originating from both the competing 
RKKY couplings to the neighboring spins and from the external magnetic field $\vec{B}$.
 At each temperature and magnetic field, the spin glass state is thus described by a distribution of effective fields 
 $P_{T,\vec{B}}(\vec{h}_i)$ over the various impurities. Each impurity is thus  treated independently of the others. The corresponding 
 magnetic dephasing can be described using the Kondo dephasing mechanism~\cite{Bauerle_05_AA,Mallet_06_AA,Saminadayar_07_AA}.
 Two different dephasing mechanisms occur depending on the strength of the effective field $|\vec{h}_i|$: for the effective "free spins" 
 for which $g \mu_B  |\vec{h}_i| \leq kT$, the dominant dephasing mechanism is the usual Kondo dephasing~\cite{Bauerle_05_AA,Mallet_06_AA,Saminadayar_07_AA} with a rate~\cite{Glazman_03_AA}
\begin{equation}
\frac{1}{\tau_{e-s}^{free}} = 
\frac{8\pi n_{imp}^{free}}{\rho(E_F)} \frac{S(S+1)}{\ln^2 T/T_K}
\label{equation1} 
\end{equation}
where
\begin{equation}
n_{imp}^{free}\simeq n_{imp}\,P_{T,B}(h=0)\,\frac{k_{B}T}{g\mu_{B}}
\label{equation2}
\end{equation}
is the concentration of these effective free spins. For the other spins (such that $g \mu_B  |\vec{h}_i| \geq kT$), the dephasing occurs \textit{via} a different virtual process mechanism with a rate
\begin{equation}
\frac{1}{\tau_{e-s}^{pol}} \simeq
\frac{n_{imp}^{pol}}{{\tau_{Kondo}(B)}}
\label{equation3} 
\end{equation}
with $\tau_{Kondo}(B) \propto B$~\cite{Glazman_03_AA}. Note that this virtual process is much less efficient than the direct process described in equation~(\ref{equation1}). 
When both are present, virtual processes are thus negligible: this corresponds to the low-field regime of figure~\ref{Fig2}. The plateau of $\Delta R$ in this
 regime hints that the probability $P_{T,B}(h=0)$ varies slowly over this magnetic field range. The change of behavior at $B\simeq 2.5\,T$ indicates the
  disappearance of the direct Kondo dephasing. Beyond this field, all impurities contribute to the dephasing through the virtual process mechanism
 (eq.~\ref{equation2}). In this regime, $\tau_\phi$ is proportional to $B$ (equation~\ref{equation3}) leading to an amplitude of the $UCFs$ scaling as
 $\Delta R \propto L_{\phi}^{3/2}\simeq B^{3/4}$, in good agreement with our experimental data (figure~{\ref{Fig2}}).

\begin{figure}[t!]
\includegraphics[width=8.5cm]{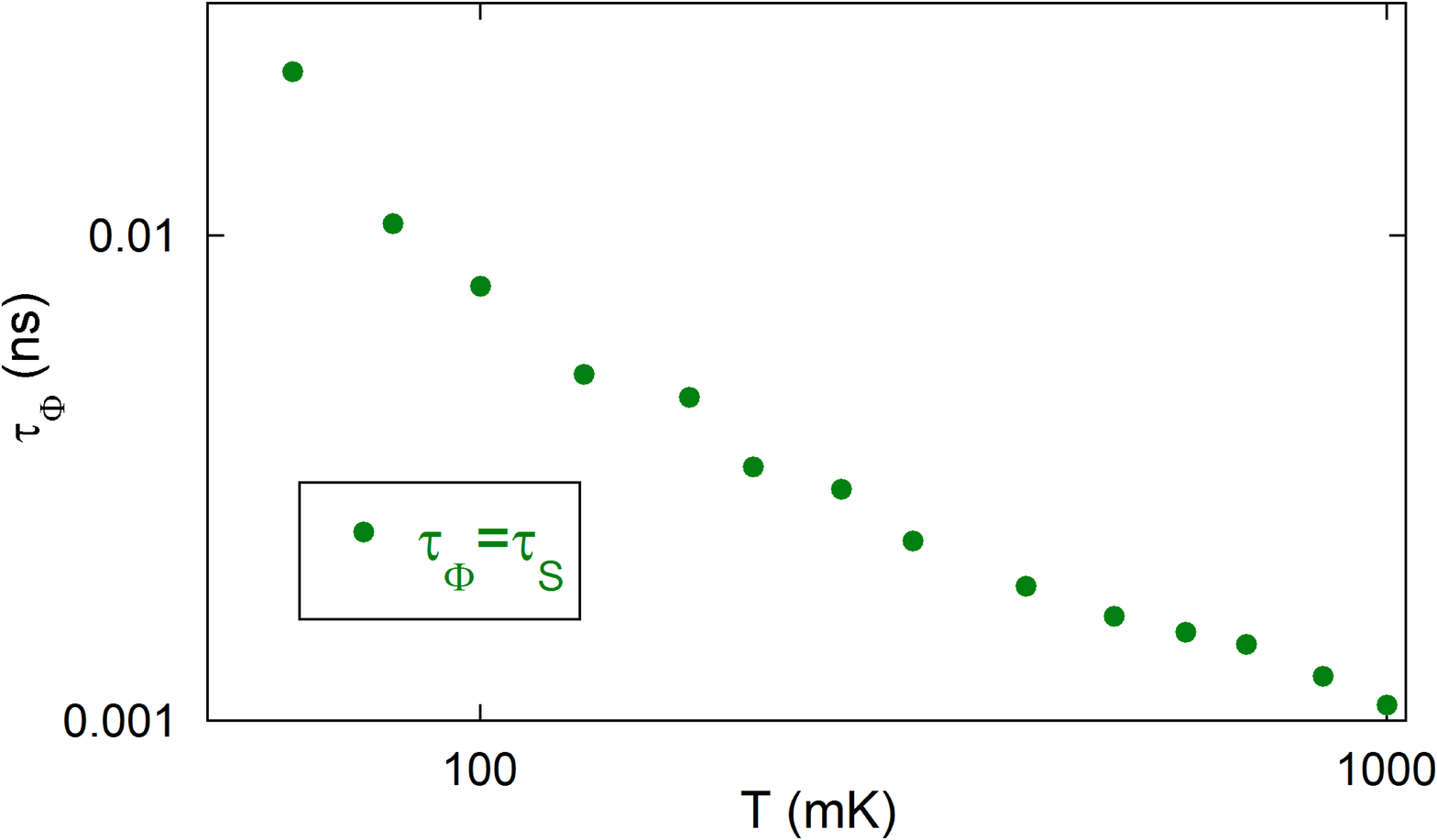}
\caption{(Color online) Decoherence rate as a function of temperature extracted from the low field curve of the figure~\ref{Fig3}.}
\label{Fig4}
\end{figure}

\begin{figure}[t!]
\centerline{
\includegraphics[width=7.5cm]{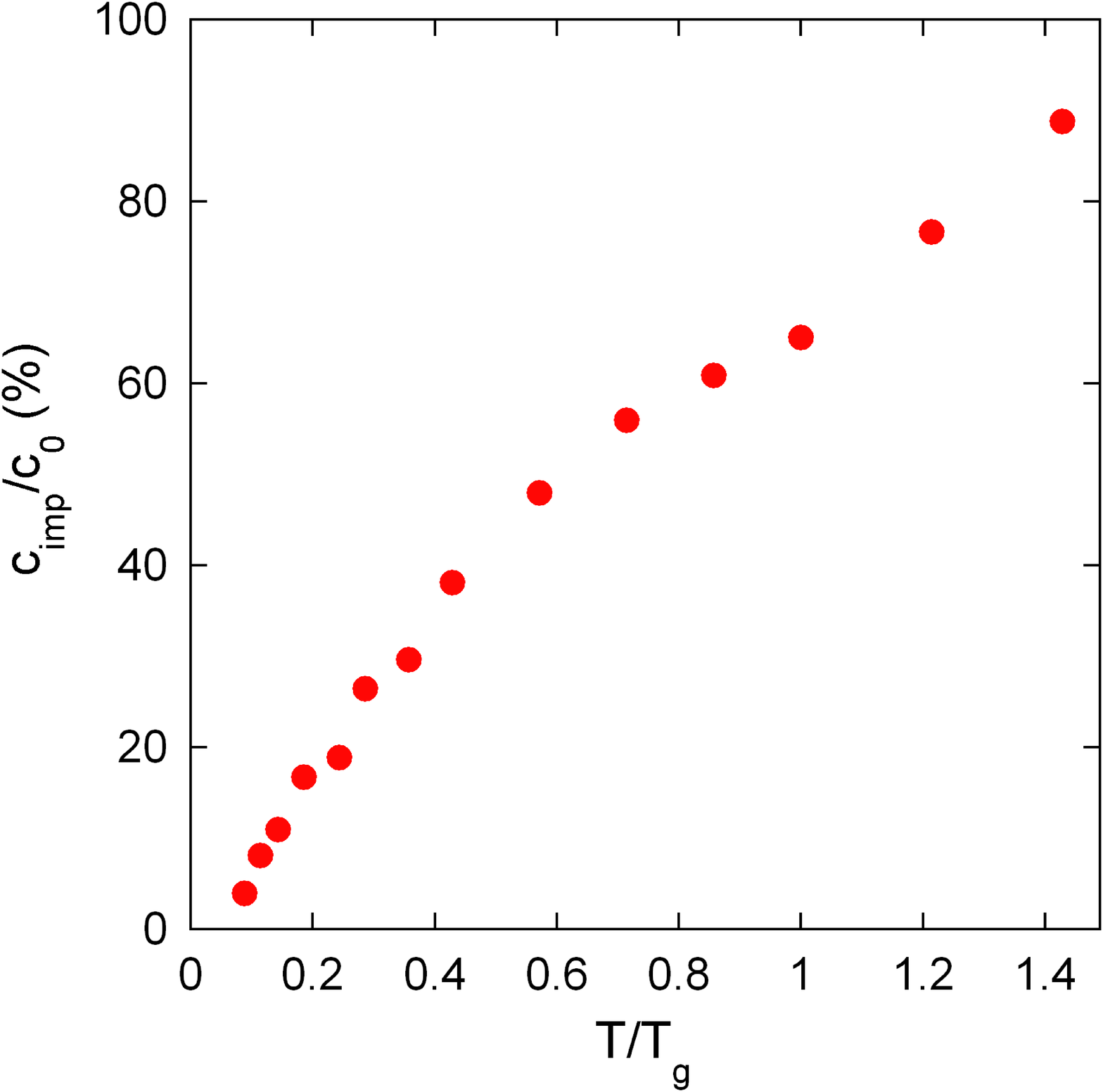}}
\caption{(Color online) Fraction of free spins (normalized to the initial concentration of magnetic impurities) as a function of 
the normalized temperature (normalized to the freezing temperature $T_{sg}$).}
\label{Fig5}
\end{figure}

On figure~\ref{Fig3}, we have plotted the amplitude of the UCFs for both high and low magnetic field, as a function of temperature. The high field ($B\geq 6\,T$)
amplitude is found to scale as $\Delta G_{\textsc{ucf}s} \propto T^{-1/2}$. This scaling is compatible with our previous analysis using 
$L_T=(\hbar D/T)^{1/2}\simeq T^{-1/2}$. To analyze quantitatively the low field regime ($B\leq2\,T$), we extract the phase coherence time $\tau_{\phi}$ from the amplitude of the \textsc{ucf}s and then 
subtract the Al'tshuler-Aronov-Khmelnitsky (\textsc{aak})~\cite{Altshuler_85_AA}: this provides the (dominant) magnetic dephasing rate. Its temperature dependence is 
depicted on figure~\ref{Fig4}. Using equation~(\ref{equation1}), and neglecting the $\ln^{2}T/T_{K}$ when $T\gg T_{K}$, we can extract the temperature
 dependance of the fraction of free spins, $n_{imp}^{free}/n_{imp}$. The result is shown on figure~\ref{Fig5}. Note, however, that above $1\,K$, the data 
 are not  reliable as the (neglected) 
electron-phonon scattering starts to play a non-negligeable role in the decoherence processes. The obtained temperature dependance is compatible with $n_{imp}^{free}/n_{imp}\propto T^{\alpha}$ with $\alpha$ close to $1$. We can extract the temperature scaling behavior of the probability $P_{T,B}(h_i=0)\propto T^{\alpha-1}$ using equation~\ref{equation2}. The experimental data are compatible with either a constant probability $(\alpha = 1)$ or a weak pseudo-gap behavior ($\alpha\geq 1$)~\cite{Boettcher:08}.

What can we infer from these measurements? In our system, both the implantation concentration as well as remanent effect 
indicate a freezing temperature of $\approx700\,mK$~\cite{Capron_11_AA}. It is commonly believed that slightly below 
$T_{sg}$ (almost) all the spins are completely frozen. Mesoscopic probes, which have so far not been exploited in the field of spin glasses, demonstrate that this is actually not the case: even at low temperature (one tenth of $T_{sg}$), almost $10\,\%$ of the spins are still free to flip by thermal activation. This shows that the spin glass transition is indeed very broad, a total 
freezing of all the spins appearing only below ${T_{sg}}/10$. Moreover, even at low temperature and below the characteristic field $B_{c}$, a finite fraction of the spins remains actually free.

To conclude, we have measured phase coherent transport in an $Ag:Mn$ spin glass doped at a level of $700\,ppm$. We 
have shown that the Universal Conductance Fluctuations are perfectly reproducible up to a field of $8\,T$, \textsl{i.e.} a 
field much larger than the characteristic field $B_{c}\approx k_{B}T_{sg}/\mu_{B}$. Moreover, we observe a strong increase 
of the amplitude of these \textsc{ucf}s above a field of $2.5\,T$; this increase in interpretated as an increase of the phase 
coherence length due to the polarization of the spins. Finally, we decribe the gradual freezing of the spins when lowering the temperature. This constitutes the first experimental probe of the fraction of free spins in a spin glass using mesoscopic transport measurements. This study paves the way for a measurement of the overlaps between microscopic spin configuration in the spin glass phase~\cite{Mezard_84_AA}.

\acknowledgments
We are indebted to H. Pothier and H. Bouchiat for the use of their Joule evaporators. We thank L. Glazman, G. Paulin, E. Orignac, B. Spivak, H. Bouchiat, \'{E}. Vincent, D. Est\`{e}ve, A. Rosch, T. Micklitz, R. Whitney and H. Alloul for fruitful discussions. This work 
has been supported by the French National Agency (\textsc{anr}) in the frame of its \textquotedblleft Programmes Blancs
\textquotedblright~(\textsc{MesoGlass} project).

\end{document}